\documentclass{appolb}
\usepackage{epsfig,wrapfig}

\begin{document}

\title{Large Deformation Effects in the N~=~Z \\
$^{44}$Ti Compound Nucleus
\thanks{Presented at the 10th INTERNATIONAL CONFERENCE ON NUCLEAR REACTION 
\\MECHANISMS Varenna, Italy, June 9-13, 2003.}\\
%% you can use '\\' to break lines
}
\author{P. Papka, C. Beck, F. Haas, V. Rauch, M. Rousseau, P. Bednarczyk, S.
Courtin, O. Dorvaux, K. Eddahbi, K. Kezzar, I. Piqueras, J. Robin, A.
S\`{a}nchez i Zafra, S. Thummerer\\
\address{Institut de Recherches Subatomiques, UMR7500,
CNRS-IN${2}$P${3}$/Universit\'{e} Louis Pasteur, 23 rue du Loess, B.P. 28,
F-67037 Strasbourg Cedex 2, France}\\
%\address{$^{2}$ GSI}\\
%\address{$^{3}$ HMI}\\
\vspace{0.3cm}
A. Hachem and E. Martin\\
\address{Universit\'{e} de Nice-Sophia Antipolis, F-06108 Nice, France}\\
\vspace{0.3cm}
N. Redon, B. Ross\'{e}, O. Stezowski, A. Pr\'{e}vost 
\address{IPN Lyon, CNRS-IN${2}$P${3}$/Universit\'{e} Lyon-1, F-69622 
Villeurbanne Cedex, France}\\
\vspace{0.3cm}
A.H. Wuosmaa
\address{Physics Division, Argonne National Laboratory, Argonne, Illinois
60439}
}

\maketitle 
\begin{abstract}

The N~=~Z $^{44}$Ti$^{*}$ nucleus has been populated in Fusion Evaporation
process at very high excitation energies and angular momenta using two
entrance channels with different mass-asymmetry. The deformation effects in
the rapidly rotating nuclei have been investigated through the energy
distribution of the $\alpha$-particle combined to statistical-model
calculations. In the case of low-multiplicity events, the ratio between first
particle emitted has been measured and shows significant disagreement with the
predictions of the statistical-model. This may explain The large discrepancies
observed in proton energy spectra measured in previous experiments performed
in the same mass region.

\noindent

\end{abstract}

  
\section{Introduction}

Signatures of deformation effects in hot rotating nuclei, produced in the
Fusion Evaporation (FE) process, have been searched for in several
experiments [1-6] using LCP emission. The level density in nuclei
increases exponentially with excitation energy and at a few tens of MeV
above the yrast line a continuum regime is reached. Consequently, LCP
energy spectra have typical Maxwellian shapes resulting from the combined
effects of the Coulomb barrier and level density. These spectra can be
described in the framework of the statistical-model which in the present
work has been exploited through CACARIZO, the Monte Carlo version of
CASCADE \cite{Puhl}. 

Up to now the main difficulty encountered in such a description has been the
lack of knowledge of the sequence in which the particles are emitted. Indeed,
depending on its excitation energy, the Compound Nucleus (CN) is able to 
evaporate several light particles which compose the decay chain but their
order can generally not be determined experimentally. At each step of the
decay a neutron, a proton or an $\alpha$-particle can be emitted  depending on
properties peculiar to each intermediate residual nuclei. There are thus a
large number of decay paths which connect the CN to the final Evaporation
Residues (ER). 

The in-plane LCP have been measured in coincidence with ER, and the data
presented in this paper have been obtained with the  multidetector array ICARE
at the VIVITRON tandem facility of the IReS (Strasbourg). The reactions
$^{16}$O on $^{28}$Si at E$_{lab}$~=~76, 96 and 112~MeV and $^{32}$S on
$^{12}$C at E$_{lab}$~=~180 and 225~MeV bombarding energies have been used to
populate the $^{44}$Ti CN at three different excitation energies between
E$^{*}$~=~60 and 82~MeV.

\section{Physics case}

%From the point of view of theoritical studies, the $^{44}$Ti is a good
%candidate to SD and HD which have never been measured \cite{qui?,qui?}. 

In reference \cite{Roye} the shape evolution of hot rotating nuclei has been
calculated using a generalized liquid drop model and a two center shell model
to describe the CN entrance channel. Concerning the $^{44}$Ti CN,

\begin{wrapfigure}[31]{r}[0.0cm]{8.0cm}
\includegraphics[width=8cm,height=12cm]{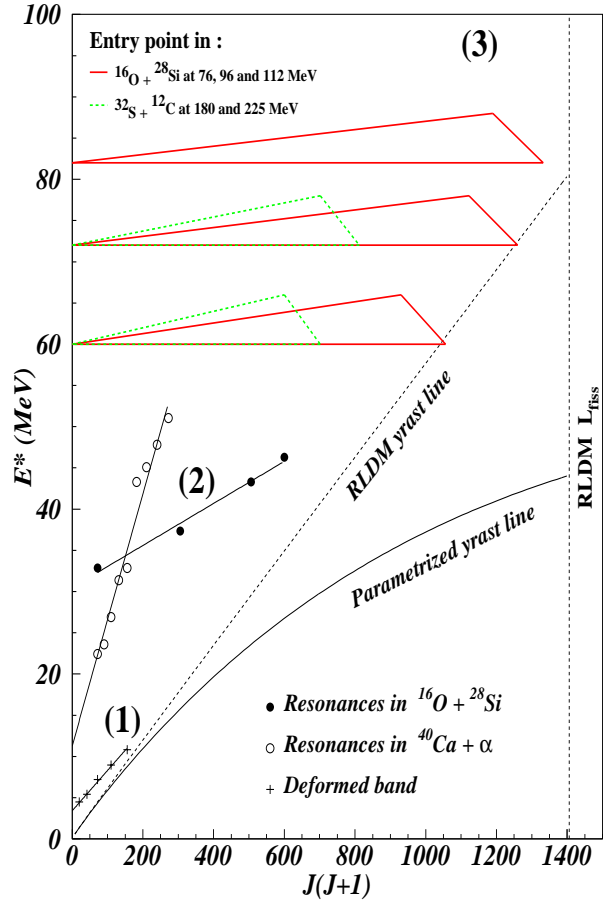}
\parbox{80mm}{
\caption{\label{region} Rotational bands in $^{44}$Ti \cite{Lear,Barr,Delb}
and excitation energy regions populated in $^{16}$O on $^{28}$Si and $^{32}$S on $^{12}$C reactions}}
\end{wrapfigure}

\noindent these calculations show that minima in the potential energy exist
not only at low excitation energy but also at higher energies in a region of
dynamical deformation and angular momenta L close to the critical angular
momentum (L$_{cr}$) above which fission occurs. 

The $\alpha$-like nucleus $^{44}$Ti lies close to the doubly magic $^{40}$Ca
nucleus in a region where spherical and deformed states coexist. In {\rm Fig.
\ref{region}}, the deformation occurence in $^{44}$Ti is displayed through the
existence of rotational bands of different origins. In this mass region,
Superdeformation at relatively low spin (J $<$ 20) has been found in $^{36}$Ar
\cite{Sven} and $^{40}$Ca \cite{Ideg} for spins up to J~=~16$^{+}$). For
$^{44}$Ti (region (1) of Fig. \ref{region}), a deformed band has been observed
up to J~=~12$^{+}$ \cite{Lear}. At higher excitation energies
(E$^{*}$~$>$~20~MeV), the existence of the so-called quasi-molecular
resonances is well known for $\alpha$-like from $^{24}$Mg to $^{56}$Ni and has
been observed essentially in elastic and inelastic scattering reactions
leading to these nuclei as composite systems. In region (2) of Fig.
\ref{region}, the alignement of E$^{*}$ versus J(J+1) is interpreted as
rotational bands of nuclear molecules $^{40}$Ca + $\alpha$ and $^{16}$O +
$^{28}$Si, the deformation being larger for the more mass-symetric systems.
These resonances are located well above $^{44}$Ti yrast line and are observed
up to L~=~24 $\hbar$. At even higher excitation energies and angular momenta
corresponding to region (3) of Fig. \ref{region}, there is a possibility to
observe the Jacobi transitions. Indeed, for the neighbouring nucleus
$^{46}$Ti, measurements of the Giant-Dipole-Resonance (GDR) indicate the 
transition (as a function of decreasing L) of the dynamical deformation from
prolate to triaxial and to oblate shape \cite{Adam}. This transition region is
expected in the angular momentum range L~=~26 to 30 $\hbar$. In region (3)
previous measurements from LCP spectroscopy have been performed to study the
dynamical deformation in various nuclei like $^{40}$Ca \cite{Forn,Rous02},
$^{56}$Ni \cite{Rous02,Bhat}, $^{59}$Cu \cite{Vies,Huiz}. The main parameters
governing the decomposition of the excited CN are L$_{cr}$ (deduced from the
complete fusion cross section using the sharp cut-off model), the CN
excitation energy and the yrast line position which defines the level density.
The study of deformation effects is essentially based on the comparison
between the measured and calculated $\alpha$-particle spectra which are the
most sensitive to deformation. Other observables like angular correlations,
mass distribution or proton energy spectra allow us to test the consistency of
the calculations.

\subsection{Deformation effects in LCP emission}

\begin{figure}[hp!]  \begin{center}     
\epsfig{figure=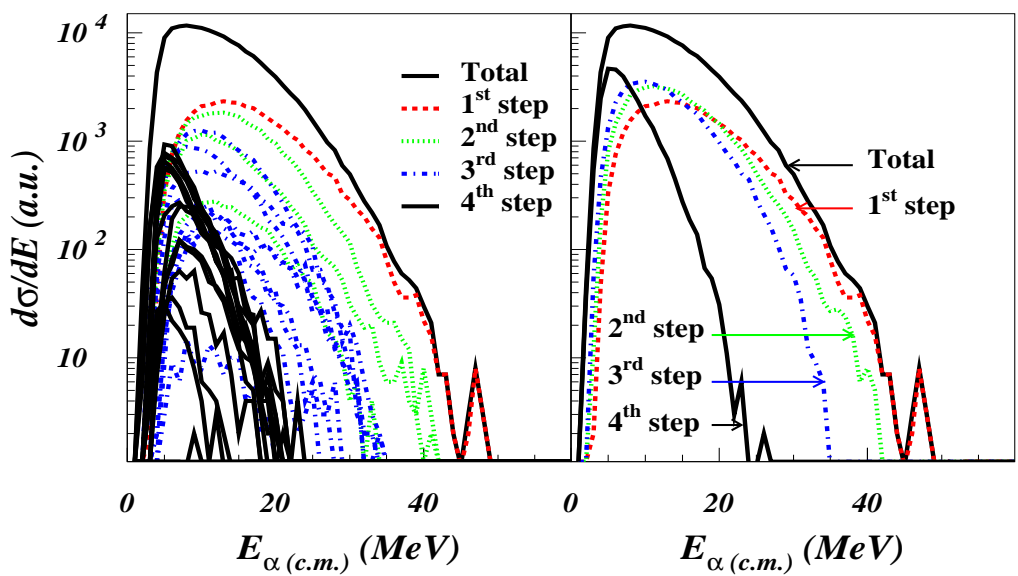,width=14cm,height=8cm} 
\caption{\label{spa} Cumulative energy spectra of $\alpha$-particle (in the
c.m.) calculated with deformability parameters $\delta$$_{1}$, $\delta$$_{2}$
\cite{Paul} corresponding to a deformation of 2:1 in the CN with L$_{cr}$~=~34
$\hbar$. In both parts of the figure, the total spectrum (full line) is the
cumulative spectrum summed over all contributions. On the left side, the total
spectrum is decomposed in each channel contribution considering their
branching ratios (dashed lines except for the 4$^{th}$ step). On the right
side, the cumulative spectrum is decomposed in the summed step contribution
when the $\alpha$-particle is emitted in the n$^{th}$ step.}     \end{center}
\end{figure}

If the yrast line of these hot rotating nuclei is calculated with the Rotating
Liquid Drop Model (RLDM), the $\alpha$-particle differential cross section is
systematically overestimated in the high-energy part of the spectra. This
energy "shift" has been interpreted as arising from the yrast line lowering
due to the change of the CN moment of inertia in the region of high angular
momenta \cite{Chou}.  The yrast line position, i.e. the effective moment of
inertia, is thus adjusted in order to reproduce the $\alpha$-particle energy
spectra. The CN is considered as a rigid sphere at L~=~0 and as an elongated
ellipsoid with a maximum axis ratio value at L~=~L$_{cr}$. The adopted yrast
lines are represented in Fig. \ref{region}: the yrast line from RLDM and the
parametrized one as calculated with the two adjustable parameters
$\delta$$_{1}$ and $\delta$$_{2}$ (see \cite{Rous02,Bhat} for more details)
obtained from a study of the $^{16}$O on $^{28}$Si fusion reaction
\cite{Paul}. The proton energy distributions are only slightly sensitive to 
deformation as the proton takes away from the CN or intermediate residual much
less energy and angular momentum than the $\alpha$-particle. 

The first chance $\alpha$-particle spectrum corresponds to the decay of
$^{44}$Ti$^{*}$ to $^{40}$Ca$^{*}$. Then, the $\alpha$-particle at the second
step can be emitted from the three nuclei: $^{40}$Ca$^{*}$, $^{43}$Sc$^{*}$
and $^{43}$Ti$^{*}$. The second step spectrum in Fig. \ref{spa} (right side)
is thus the cumulative spectrum of the three possible decay channels and the
n$^{th}$ step is the sum of 3$^{n-1}$ possible contributions. It has
previously been shown \cite{Paul} that the high-energy part of the LCP energy
spectra is determined by the first emitted particle. This is shown in the
calculated $\alpha$-particle energy spectra represented in Fig. \ref{spa}. On
both parts of the Fig. \ref{region}, the upper full line is the cumulative
spectra which is the sum of all contributions. Fig. \ref{spa} shows that the
tail (high-energy part) of the first step $\alpha$-particle represents almost
the total spectrum which justify to fit the experimental data by varying the
deformation parameters. This is not the case in the vicinity of the Coulomb
barrier which results from a large number of contributions. In this region,
not only the spectrum shape but also the branching ratio of the different
contributions has to be reproduced. From Fig. \ref{spa} it appears that for
evaporation chains with a small number of emitted particles  contributions to
the cumulative spectra are limited. A comparaison between experimental and
calculated energy spectra is then possible over the total energy range. It is
also possible to select chains with different kinetic energy distributions
from the E$_{LCP}$-E$_{ER}$ bidimensional spectra which will be discussed in a
forthcoming section. It is also worthwhile to note that the observed
branching ratios are in disagreement with the statistical-model calculations.

\section{Experimental techniques and chosen reactions} 

\begin{figure}[hp!]  \begin{center}     
\epsfig{figure=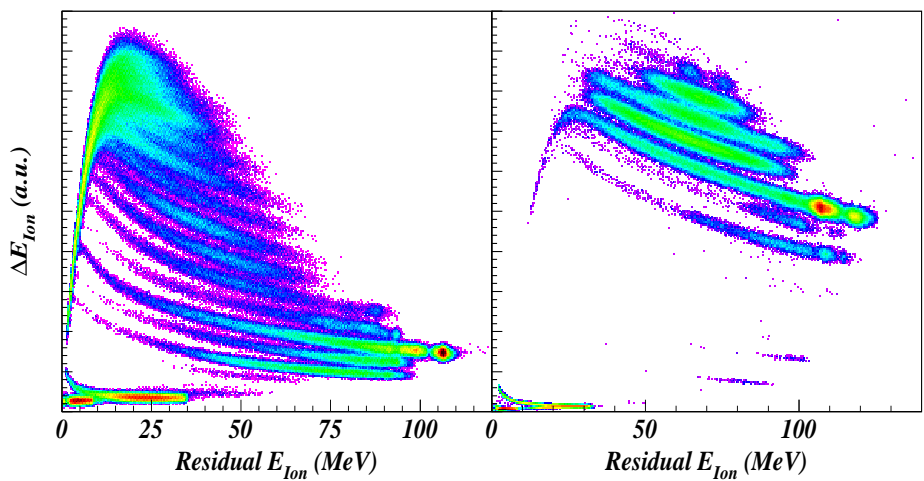,width=14cm,height=8cm} 
\caption{\label{bidim} E-$\Delta$E bidimensional spectra from IC detectors. On
the left side the $^{16}$O on $^{28}$Si reaction at E$_{lab}$~=~112~MeV with
HI detected at $\theta$$_{lab}$~=~-20$^{\circ}$ and on the right side the
$^{32}$S on $^{12}$C reaction at E$_{lab}$~=~140~MeV with HI detector placed
at $\theta$$_{lab}$~=~-10$^{\circ}$.}  \end{center} \end{figure}

The experiments were carried out using the multidetector array ICARE
\cite{Rous02} which is a combination of Heavy Ion (HI) and LCP detectors. The
HI and LCP identifications, energy and angular distributions required for
these measurements have been achieved with various telescopes composed of
Si(SB) detectors, CsI(Tl) scintillators and Ionization Chambers (IC). The IC
($\Delta$E from the gas cell and E from 500 $\mu$m Si(SB)) were placed at
forward angles to detect ER from $\theta$$_{lab}$~=~-10$^{\circ}$ to
-30$^{\circ}$ with an angular opening of $\delta$$\theta$~=~3$^{\circ}$. The
HI identification has been obtained with the E-$\Delta$E technique using a 4.8
cm length gas cell filled with isobutane at pressure of 15 and 80 Torr for the
direct kinematic reaction $^{16}$O on $^{28}$Si and inverse kinematic reaction
$^{32}$S on $^{12}$C, respectively. Fig. \ref{bidim} displays the E-$\Delta$E
HI spectra obtained in both reactions: for the $^{16}$O on $^{28}$Si reaction
(left), a threshold energy of $\approx$~20~MeV for ER has to be taken into
account in the analysis. For the $^{32}$S on $^{12}$C reaction (right) ER are
resolved up to Z~=~20 without significant detection thresholds. The LCP double
telescopes (40 $\mu$m Si(SB) + 2cm CsI(Tl)) with a
$\delta$$\theta$~=~7.5$^{\circ}$ angular opening were placed at angles from
35$^{\circ}$ to 130$^{\circ}$ to cover the LCP distribution for the $^{16}$O
on $^{28}$Si reaction. For the $^{32}$S on $^{12}$C reaction, triple
telescopes (40 $\mu$m Si(SB) + 300 $\mu$m Si(SB) + 2cm CsI(Tl)) were placed at
very forward angles ($\theta$$_{lab}$~=~30$^{\circ}$ to 40$^{\circ}$), in
order to detect with good accuracy the high-energy protons. For angles larger
than 45$^{\circ}$, double telescopes were used to detect LCP with an angular
opening of $\delta$$\theta$~=~4$^{\circ}$. In both experiments, the LCP and HI
detectors are placed with angular intervals of $\Delta$$\theta$~=~5$^{\circ}$.
The LCP discrimination was achieved using both E-$\Delta$E and time-of-flight
techniques allowing energy thresholds to be lower than 100 keV.

The $^{16}$O on $^{28}$Si reaction has been chosen to populate the $^{44}$Ti CN at
the highest angular momenta. The adopted bombarding energies of E$_{lab}$~=~76, 96
and 112~MeV correspond to L$_{cr}$~=~31, 34 and 35 $\hbar$ and E$^{*}$~=~60, 72 and
82~MeV, respectively. These values are close to the limit of L$_{fiss}$ predicted
by the RLDM. In the $^{32}$S on $^{12}$C reaction, the CN $^{44}$Ti was populated
at E$^{*}$~=~60 and 72~MeV, the same excitation energies than for the $^{16}$O on
$^{28}$Si reaction but with much lower L$_{cr}$ values (25 and 27 $\hbar$) due to
the bigger mass-asymmetry of the entrance channel. Fusion cross sections are
reported in the literature at similar CN excitation energy, populated with the same
reactions \cite{Zing,Pirr}, which allow us to extract experimental L$_{cr}$ values
using the sharp cut-off model. 

\section{Results}

\subsection{proton spectral shapes}

Despite the fact that proton energy distributions were well reproduced by the
statistical model in previous investigations \cite{Bhat}, they are completely
in disagreement for both reactions. A number of possible reasons were
suggested invoking alternate reaction mechanisms as preequilibrium or
secondary particles emitted from the  projectile-like or the target-like
nuclei. But these hypothesis are not justified as the ER and their associated
LCP detected in coincidence are consistent with FE process. 

\begin{figure}[hp!] \begin{center}     
\epsfig{figure=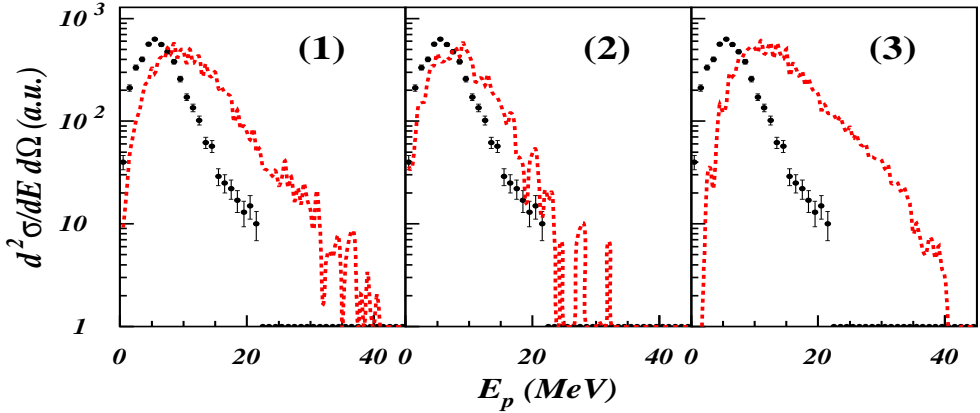,width=13cm,height=6cm}  \caption{
\label{p3929} Experimental proton energy spectra (full line), detected at
$\theta$$_{lab}$~=~60$^{\circ}$ in coincidence with Z~=~19 ER deviated at
$\theta$$_{lab}$~=~-10$^{\circ}$, measured in $^{32}$S on $^{12}$C reaction at
E$_{lab}$~=~180~MeV. The calculations surimposed (dotted lines) result from no
restrictions in cascades (1) and $\alpha$-particle (2) or proton (3) are
forced in the first step emission.}  \end{center} \end{figure}

It is crucial to know the origin of the observed proton as their energy
spectra are strongly dependent on the available energy of the emitter. As
shown in reference \cite{Sara}, where proton energy have been measured in
coincidence with discrete $\gamma$ transitions, large shifts in proton energy
spectra in the decay of $^{86}$Zr$^{*}$ CN was interpreted as near-yrast
stretched proton emission. In order to investigate the effect of the branching
ratios, the calculations have been carried out with conditions on the
cascades. In Fig. \ref{p3929}, the experimental proton energy spectrum (full
line), measured at $\theta$$_{lab}$~=~60$^{\circ}$ in coincidence with Z~=~19
ER detected at $\theta$$_{lab}$~=~-10$^{\circ}$, is compared with the
calculations (dashed lines). In these events only two charge units are not
detected: one $\alpha$-particle in the {\it 1p1$\alpha$xn} channels or two
protons in the {\it 3pxn} channels. In the calculated spectrum (1) there is no
restrictions which shows the disagreement observed in every experiments and,
in a worst case, spectrum (3) is a calculation restricted to a proton emitted
at the first step of the cascade. Indeed, the slope of the experimental proton
energy spectra indicates a lower temperature of the emitter as compared with
the predictions. Then, spectrum (2) is calculated considering the cascades
starting with the emission of one $\alpha$-particle, thus, the remaining
$^{40}$Ca$^{*}$ evaporates a light particle with a lower nuclear temperature.
The disagreement is still remaining in (2) but blocking the first proton
particle emission gives a better agreement especially in the high-energy
slope. To go ahead, the following steps of the cascades have to be better
studied to reproduce the complete spectra, particularly in the Coulomb barrier
region.

\subsection{$\alpha$-particle spectral shapes} 

\begin{figure}[hp!]  \begin{center}	
\epsfig{figure=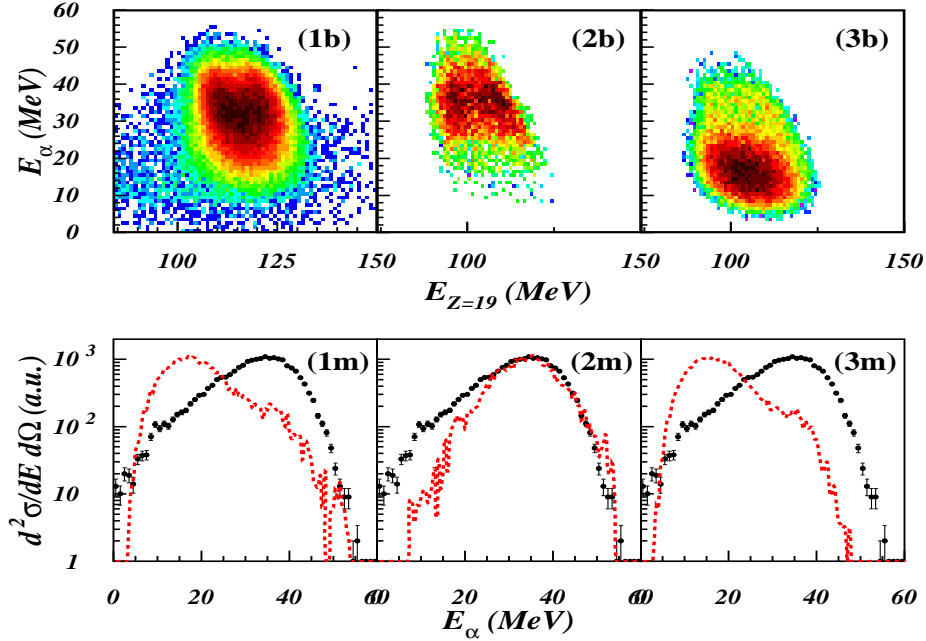,width=12cm,height=9cm}  \caption{
\label{a3929}(1b) is the experimental E$_{\alpha}$-E$_{Z=19}$ bidimensional
spectrum measured ($\theta$$_{\alpha}$~=~60$^{\circ}$,
$\theta$$_{HI}$~=~-10$^{\circ}$) in $^{32}$S on $^{12}$C reaction at
E$_{lab}$~=~180~MeV, (2b) and (3b) are calculated. (2b) results from the
selection of cascades with $\alpha$-particle emitted in the first step and
(3b) the same selection dedicated to protons. The projections (1m,2m,3m)
correspond to the bidimensional spectra above with experimental (full line)
and calculated (dotted line) $\alpha$-particle energy spectra. (1m) is the
standart calculation with no restrictions on the cascades.}   \end{center} 
\end{figure}

In the $^{16}$O on $^{28}$Si reaction the fitting of the $\alpha$-particle
energy spectra indicates strong deformation effects with an axis ratio of
$\approx$~2:1 consistent with the Superdeformation in the CN at L$_{cr}$
\cite{Paul}. In the $^{32}$S on $^{12}$C experiment at 180~MeV, the same
$\delta$$_{1}$ and $\delta$$_{2}$ deformability parameters extracted from the
latter measurement, have been used to perform the calculations displayed in
Fig. \ref{a3929}.  The disagreement in the $\alpha$-particle energy spectrum
(1m) is connected to the reason invoked in the previous paragraph. The Z~=~19
ER, considered for proton in Fig. \ref{p3929}, is measured in coincidence with
an $\alpha$-particle in the E$_{\alpha}$-E$_{Z=19}$ bidimensional spectrum
(1b) where {\it 1pxn} are missing in the measurement. This spectrum can be
splitted in the calculated ones (2b) and (3b). The former is the contribution
of the particular {\it $\alpha$p} cascade corresponding to the upper part of
the E$_{\alpha}$-E$_{Z=19}$ experimental distribution and, on the
corresponding projections (2m), the bump centered at E$_{lab}$~=~35~MeV is
well reproduced. The difference between data and calculations, in the
low-energy part of (2m), corresponds to the cascades starting with a proton
represented in spectra (3m) and (3b). The latter contribution represents few
percent of the total  which is much weaker than the predicted ratio in (1m).
This point demonstrates that the de-excitation through the $\alpha$-particle
emission is favoured against proton evaporation.

\begin{figure}[hp!]  \begin{center}    
\epsfig{figure=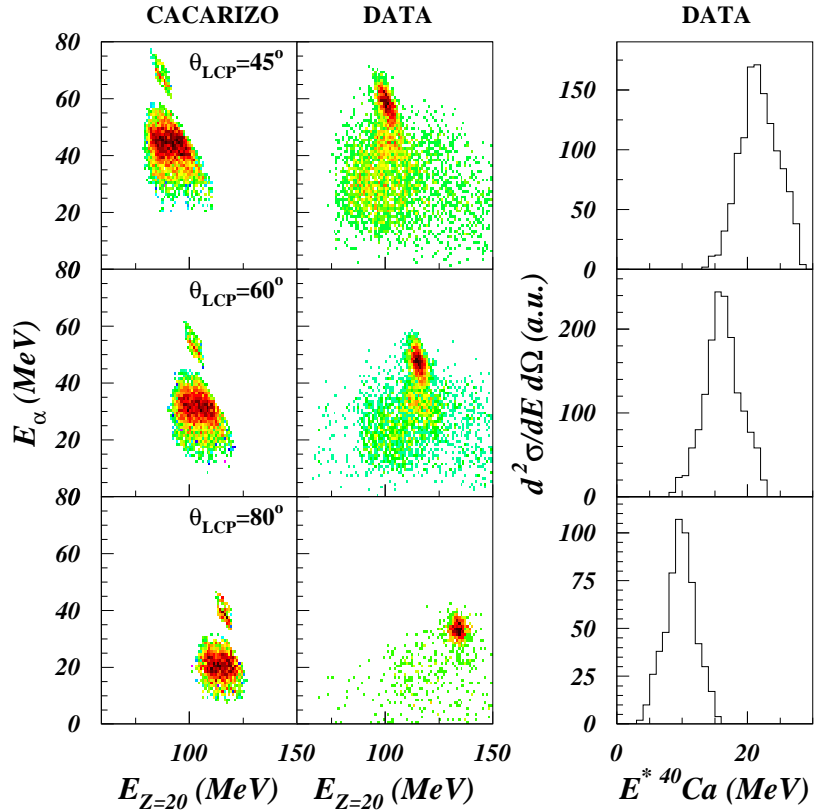,width=11cm,height=11cm}  \caption{
\label{exca}  E$_{\alpha}$-E$_{Z=20}$ bidimensional spectra: first row is the
result of  CACARIZO calculation and the second one is the experimental data for
ER detected at $\theta$$_{lab}$~=~-10$^{\circ}$ and LCP at $\theta$$_{lab}$=
45$^{\circ}$, 60$^{\circ}$ and 80$^{\circ}$. The monodimensional spectra
represent the excitation energy in $^{40}$Ca deduced from the kinetic energy
measurement in the corresponding E$_{\alpha}$-E$_{Z=20}$  spectra.}
\end{center} \end{figure}

In the $^{32}$S on $^{12}$C experiment the single particle emission has been
measured in the FE process displayed on the Fig. \ref{exca}. The structures
visible in the experimental data seem to arise from a direct process as
previously evidenced in the $^{12}$C($^{28}$Si,$^{32}$S)$^{8}$Be
$\alpha$-particle transfer reaction  \cite{Rous02}. In the experimental
bidimensional E$_{\alpha}$-E$_{Z=20}$ spectra (second column), the narrow peak
represents the $\alpha$-particle detected in coincidence with the $^{40}$Ca.
This evaporation channel is well reproduced by the statistical-model
calculations (first column). The wider component with $\alpha$-particle energy
distribution below the one-$\alpha$ region corresponds to the {\it
1$\alpha$xn} decay channels. It is interesting to note the large excess of the
neutron evaporation yield predicted in the calculations. As shown in Table
\ref{ratio} the neutron emission leading to Z~=~20 ER is overestimated with a
factor of $\approx$~5 in comparison with the experimental values. 

The comparison of the ratio between {\it 1$\alpha$} and {\it 2pxn} cascades
reported in Table \ref{ratio} show the underestimation of $\alpha$-particle
emission in the standart calculation with more than one order of magnitude.
The two-body kinematics, by means of $\theta$$_{\alpha}$-$\theta$$_{Z=20}$
angle combination, imposes the excitation energy in the remaining
$^{40}$Ca$^{*}$, shown in the last column of the Fig. \ref{exca}, deduced from
the kinetic energies measured at the indicated angles. At the most forward
angle, the detection of the bound $^{40}$Ca indicates that it is populated at
excitation energies in the region of the highest Superdeformed bands
transitions measured in \cite{Ideg}. The energy spread is due to the angular
opening of the detectors as the $\alpha$-particle energy and excitation energy
in $^{40}$Ca$^{*}$ have a strong angular dependence. However, the centroids
give the mean energies at the considered angles and the resolution is as well
reproduced in the calculations.

\begin{table}[hp!]
\begin{center}
%\hspace{-2cm}
{\label{ratio}}
\begin{tabular}{|c|c|c|c|}
\hline
Detection angle   	& measured {\it 1$\alpha$/2pxn} & calculated {\it 1$\alpha$/2pxn} & ratio \\ 
\hline
45$^{\circ}$	    	&	    4.7		   &	    0.22		&      	21.2		    \\
\hline
60$^{\circ}$	    	&	    4.1		   &	    0.16		&	25.6		    \\
\hline
80$^{\circ}$	    	&	    3.9		   &	    0.38		&	10.2		    \\
\hline
\hline
Detection angle   	& measured {\it 1$\alpha$/1$\alpha$xn} & calculated {\it 1$\alpha$/1$\alpha$xn} & ratio \\ 
\hline
45$^{\circ}$	    	&	    0.27	   &	    0.057	     	&      	4.7		    \\
\hline
60$^{\circ}$	    	&	    0.40	   &	    0.059		&	6.8		    \\
\hline
80$^{\circ}$	    	&	    0.61	   &	    0.170		&	3.6		    \\
\hline
\end{tabular}

       \caption{{\it 1$\alpha$/1$\alpha$xn} and {\it 1$\alpha$/2pxn} ratio
leading to Z~=~20 ER from both  experimental data and calculations in $^{32}$S
on $^{12}$C reaction at E$_{lab}$~=~180~MeV. $\alpha$-particles are detected
at angles indicated in the table in coincidence with HI at
$\theta$$_{lab}$~=~-10$^{\circ}$.}

\renewcommand{\baselinestretch}{1.}
\end{center}
\end{table}

\section{Conclusions} 

According to the high energy slopes of the $\alpha$-particle spectra, measured
in the $^{16}$O on $^{28}$Si reaction, the temperature in the emitters is well
reproduced. Thus, the discrepancies in the proton energy spectra have been
understood as a misinterpretation in the branching ratio in particle emission
from highly excited CN. Indeed, the experimental results compared with
calculations give an indication concerning the possible order of the emitted
particles and indicate  that the $\alpha$-particles are more favourably
emitted at the beginning of the de-excitation chain. Consequently, the ratio
between particles emitted at each step of the cascade should be considered
more precisely to go further in the analysis of LCP energy spectra. In this
paper the study of low-multiplicity events allows the complete measurements of
the reaction. Concerning the one-$\alpha$ evaporation channel, the kinematic
measurement gives the opportunity to select particular excitation energy
regions in the residual nuclei. For the search for Superdeformed or
Hyperdeformed bands $\gamma$-ray spectroscopy combined with particle detection
offer the possibility to select highly excited bound states in the studied
nuclei such as the N~=~Z doubly magic nucleus $^{40}$Ca.

\vspace{1cm}
{\bf Acknowlegements:} 
\vspace{0.3cm}

The authors wish to thank the staff of the VIVITRON for providing us with good $^{16}$O and
$^{32}$S stable beams, M.A.~Saettel for preparing the targets, and J.~Devin and C.~Fuchs for the
excellent support in carrying out these experiments. We wish also to thank N.~Rowley for valuable
discussions during the progress of this invetigations. 

\newpage

\label{biblio}

\end{document}